\documentclass[aps,prl,twocolumn,10pt]{revtex4-1}
\usepackage{graphicx,amsmath,amssymb}
\usepackage[utf8]{inputenc}
\usepackage{url}

\begin{document}

\title{Light transport and localization in two-dimensional correlated disorder}

\author{Gaurasundar M. Conley$^{1,2}$}
\author{Matteo Burresi$^{1,3}$}
\email[]{burresi@lens.unifi.it}
\author{Filippo Pratesi$^1$}
\author{Kevin Vynck,$^{1,4}$}
\altaffiliation{Current affiliation: Laboratoire Photonique, Num\'erique et Nanosciences (LP2N), UMR 5298, CNRS - IOGS - Univ. Bordeaux, Institut d'Optique d'Aquitaine, 33400 Talence, France}
\author{Diederik S. Wiersma$^{1,3}$}

\affiliation{$^1$European Laboratory for Non-linear Spectroscopy (LENS), University of Florence, Via Nello Carrara 1, 50019 Sesto Fiorentino, Italy}
\affiliation{$^2$Physics Department, University of Fribourg, Chemin du Mus\'{e}e 3, 1700 Fribourg, Switzerland}
\affiliation{$^3$National Institute of Optics (CNR-INO), Largo Fermi 6, 50125 Florence, Italy}
\affiliation{$^4$Institut Langevin, ESPCI ParisTech, CNRS, 1 rue Jussieu, 75238 Paris Cedex 05, France}

\date{\today}

\begin{abstract}
Structural correlations in disordered media are known to affect
significantly the propagation of waves. In this article, we
theoretically investigate the transport and localization of light
in 2D photonic structures with short-range
correlated disorder. The problem is tackled semi-analytically
using the Baus-Colot model for the structure factor of
correlated media and a modified independent scattering
approximation. We find that short-range correlations make
it possible to easily tune the transport mean free path by more than a
factor of 2 and the related localization length over several
orders of magnitude. This trend is confirmed by numerical
finite-difference time-domain calculations. This study therefore
shows that disorder engineering can offer fine control over
light transport and localization in planar geometries, which may
open new opportunities in both fundamental and applied photonics
research.
\end{abstract}

\maketitle

Multiple light scattering in disordered media plays a paramount
role in the study of complex natural systems (e.g., biological
tissues, porous materials, planetary
atmospheres)~\cite{Ishimaru1999} and wave phenomena (e.g., light
localization, anomalous diffusion)~\cite{Akkermans2007, Sheng2010,
Wiersma2013}. In recent years there has been a growing interest in
the use of photonic structures with \textit{controlled} disorder,
in particular within the context of mesoscopic transport
effects~\cite{Riboli2011, Garcia2012, Burresi2012, Strudley2013},
cavity quantum electrodynamics~\cite{Sapienza2010}, photon
management for energy efficiency~\cite{Vynck2012a,
burresi_two-dimensional_2013, Oskooi2012, Lin2013} and even
lab-on-chip spectroscopy~\cite{Redding2013}. Indeed structural
correlations in the positions of scatterers are known to affect
light propagation. Previous studies have shown that short-range
correlations can either diminish or enhance the scattering
strength of a disordered system~\cite{Fraden1990, Saulnier1990,
RojasOchoa2004, Reufer2007} and lead to a modulation of the
density of optical states~\cite{Yang2010}, even in biological
systems~\cite{Yin2012}. Such a modulation can be so large that a
complete photonic bandgap is expected to form, even without
long-range periodicity~\cite{Edagawa2008, Florescu2009,
Imagawa2010, Yang2010, Rechtsman2011}. The emerging concept of
``disorder engineering'' to manipulate light transport in random
media is, however, still in its infancy and little is known so far
on the occurrence of localization phenomena in correlated systems.

In this article, we theoretically investigate the transport of
light and the occurrence of localization in two-dimensional (2D)
photonic structures possessing short-range correlated disorder. A
semi-analytical model describing the wave propagation in
correlated-disordered systems allows us to investigate how key
quantities, namely the transport mean free path, the scattering
anisotropy factor and the localization length, evolve with the
degree of correlation. In particular, short-range correlations are
found to allow for the tuning of the localization length over
several orders of magnitude and thus, make it possible to go from
a quasi-extended to a strongly localized regime very easily, in
sharp contrast with three-dimensional systems, where the localized
regime is very difficult to reach
\cite{Sperling_direct_2013,Strudley2013}. This trend is confirmed
by numerical simulations.

\begin{figure}[htb]
\includegraphics[width=6cm]{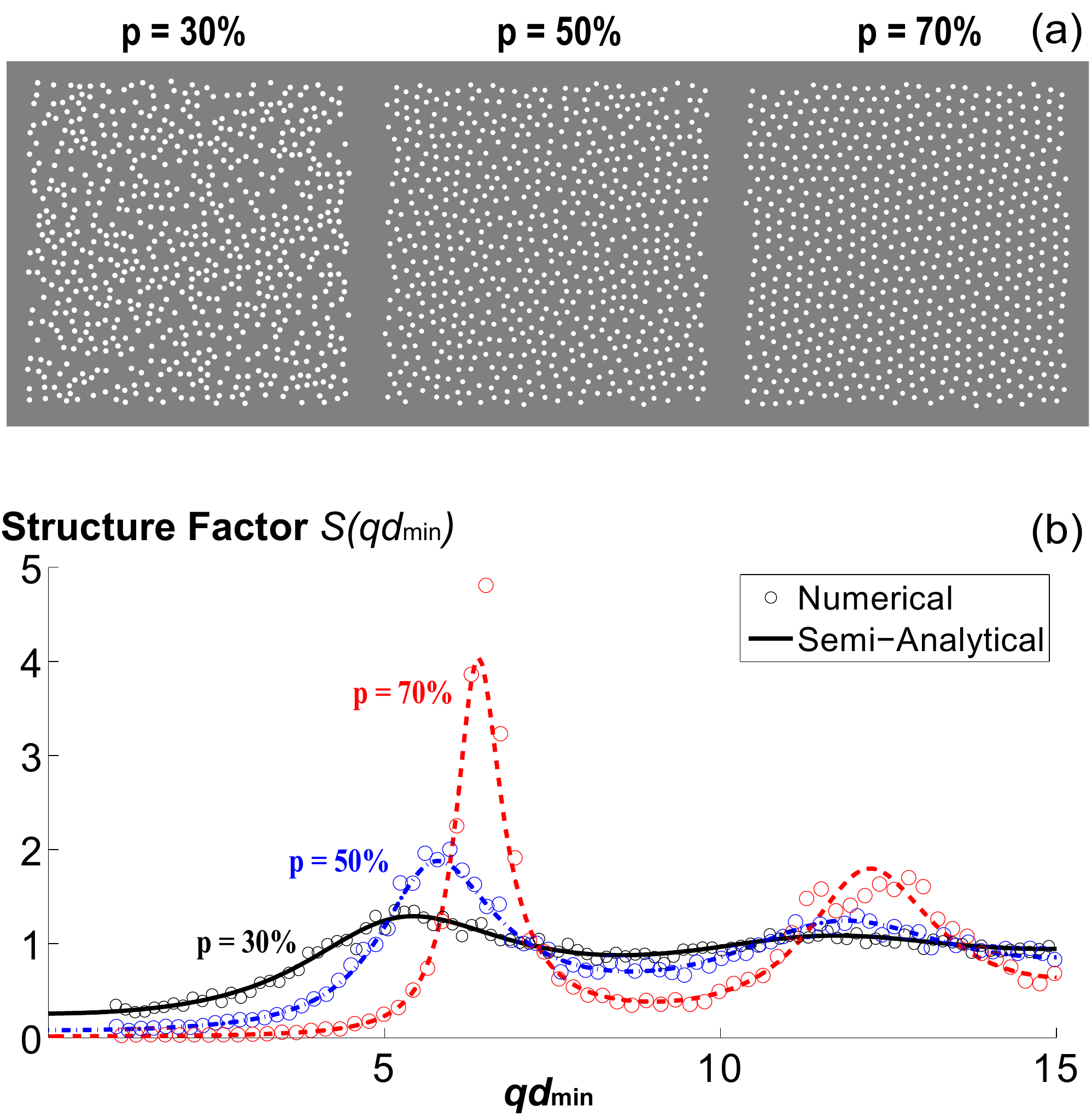}
\caption{\textbf{Structural correlations in 2D disordered media} --
(a) Disordered photonic structures with filling fraction $f=20\%$
and different degrees of correlation ($p=30$, $50$, $70\%$). (b) Structure
factors $S(qd_\text{min})$ of the correlated patterns, evaluated
numerically with Eq.~\ref{eq1} (hollow
dots) and semi-analytically using the Baus-Colot model (solid
lines).
\label{fig1}}
\end{figure}

The 2D photonic structures consist of disordered patterns of
circular air holes ($n_i=1$) with filling fraction $f=20\%$ and
diameter $\o=0.23a$, where $a$ is the period of a hexagonal
lattice of holes with the same $f$,  in a medium with refractive
index $n_o=3.5$~\cite{Yang2010, Vynck2012a}. The short-range
correlation in the disorder is controlled by imposing a minimum
distance $d_{\text{min}}$ between the centers of the holes. This
has been obtained \cite{Vynck2012a,pratesi_disordered_2013} by
generating a disordered packing of hard disks with diameter
$d_{\text{min}}$ at a packing fraction $p$, using a
freely-available code~\cite{Skoge2006} based on the
Lubachevsky-Stillinger algorithm~\cite{Lubachevsky1990}. The
centers of these disks have been used to generate the point
patterns shown in Fig.~\ref{fig1}(a) for three photonic structures
with  $p=30,50,70\%$. The degree of local order evidently
increases with increasing $p$, imposing an average distance
between adjacent holes.

Fundamentally, the existence of a typical distance between
neighboring scatterers implies a certain phase relation which,
depending on direction and wavelength of scattered waves in the
medium, gives rise to either constructive or destructive interference between
them. Following this line of reasoning, structural correlations
can be seen as a modification of the angular scattering pattern of
the individual scatterers, and be taken into account (to a first
approximation) by correcting the expression of the single
scatterer differential scattering cross-section $\mathrm{d}\sigma
/ \mathrm{d} \theta$ by the static structure factor $S(q)$
as~\cite{Fraden1990, Saulnier1990}
\begin{equation}\label{eq2}
\frac{\mathrm{d}\sigma^{\star}}{\mathrm{d}
\theta}=\frac{\mathrm{d}\sigma}{\mathrm{d} \theta} S(q),
\end{equation}
where $q=(4\pi/\lambda_e)\sin (\theta/2)$, $\theta$ is the
scattering angle and  $\lambda_e=\lambda/n_e$ is the wavelength in
a medium with effective refractive index $n_e$, which in our case
equals $2.92$ according to the two-dimensional Maxwell Garnett
mixing rule~\cite{Garnett1906,sihvola_electromagnetic_1999}.
Formally, the structure factor $S(\mathbf{q})$ is defined as
\begin{equation}\label{eq1}
S(\mathbf{q})= \frac{1}{N} \langle \sum_{i,j=1}^N
\text{e}^{-\mathrm{i}\mathbf{q} \cdot (\mathbf{r}_{i}-\mathbf{r}_j)} \rangle
\end{equation}
where $N$ is the number of scatterers, $\mathbf{r}_{i,j}$ the
position of the scatterers labelled $i$ and $j$, and $\langle ...
\rangle$ denotes ensemble average. In previous works on 2D
photonic structures with short-range correlated disorder
(statistically isotropic), $S(q)$ was calculated
\textit{numerically} from the point patterns generated by a sphere
packing protocol~\cite{Yang2010}. This approach is very time
consuming and not suited to an exhaustive study of the effect of
structural correlations on transport. An analytical expression of
the structure factor of a correlated-disordered medium is often
retrieved by making use of the well-known Percus-Yevick model
which, unfortunately, applies exclusively to systems with odd
dimensionality ($d=1,3,...$)~\cite{Torquato2001}. In contrast, we
adopt a \textit{semi-analytical} approach, based on the Baus-Colot
(BC) model for the structure factor of a fluid of hard
disks~\cite{Baus1987}, that is well-suited to systems of
dimensionality $d=2$. In Fig.~\ref{fig1}(b), we compare the
structure factor $S(qd_\text{min})$ evaluated numerically using
Eq.~\ref{eq1} for the 2D photonic structures generated above with
those obtained from the BC model, using $p$ as the only input
parameter. A very good agreement is observed, even for high $p$.
As the degree of correlation is increased, the structure factor
exhibits stronger oscillations, which indicate the emergence of a
typical distance between neighbouring scatterers.

\begin{figure}[t!]
\includegraphics[width=7cm]{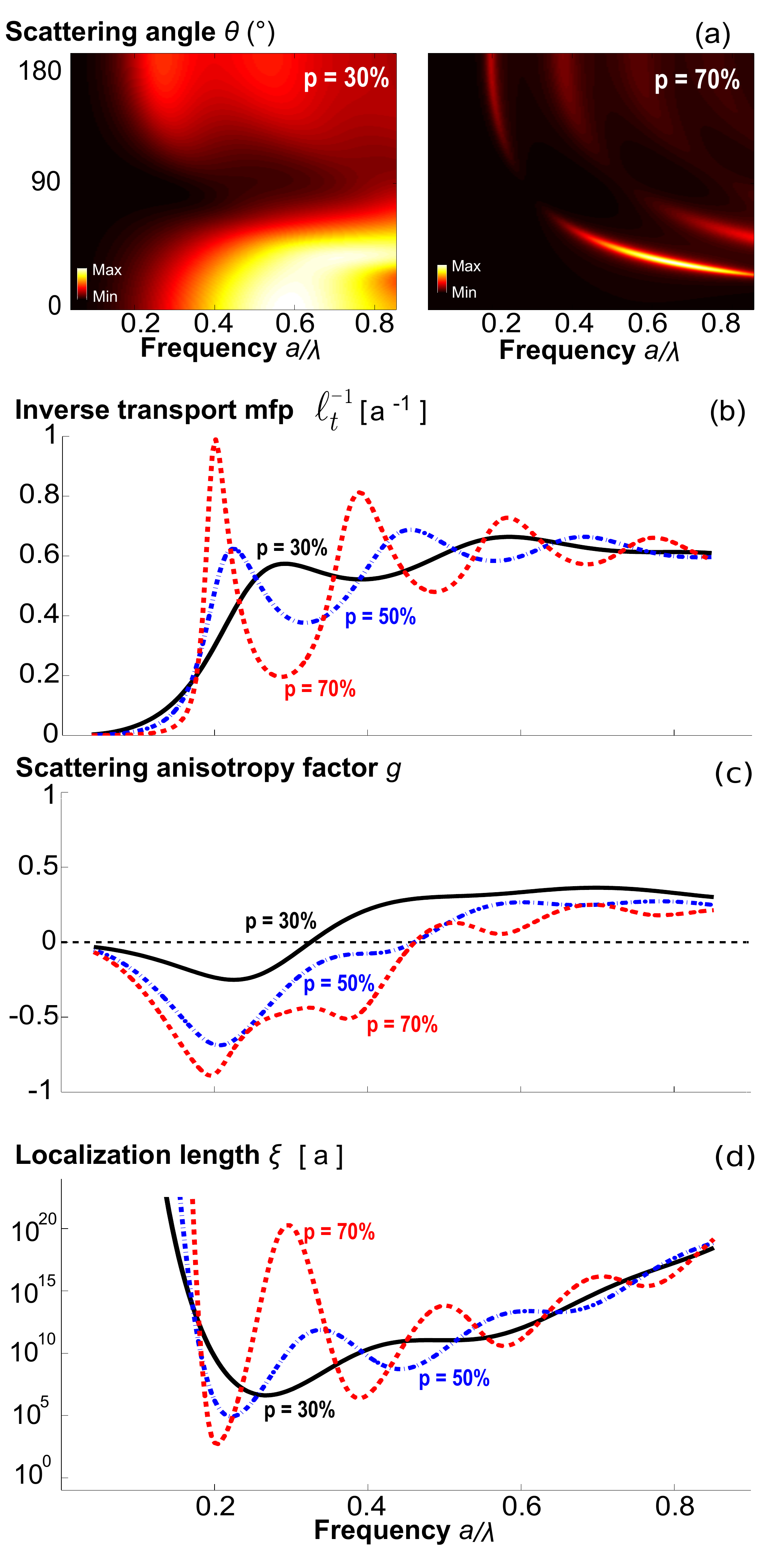}
\caption{\textbf{Modified transport due to correlations} -- (a)
Spectral and angular map of the ``effective'' differential
scattering cross-section $\mathrm{d}\sigma^{\star} / \mathrm{d}
\theta$ for weakly ($p=30\%$) and strongly ($p=70\%$) correlated
media. (b) Inverse transport mean free path $\ell_t^{-1}$, (c)
scattering anisotropy factor $g$, and (d) localization length $\xi$ for different degrees of correlation $p$.
\label{fig2}}
\end{figure}

By making use of Eq. \ref{eq2}, we calculate the angularly and
spectrally-resolved ``effective'' differential scattering
cross-section $\mathrm{d}\sigma^{\star} / \mathrm{d} \theta$ of
holes in TE-polarization (electric field in the plane)
as a function of the degree of short-range correlation
(Fig.~\ref{fig2}(a)). The single scatterer differential scattering
cross-section $\mathrm{d} \sigma / \mathrm{d} \theta$ was
calculated from Mie theory for circular
cylinders~\cite{Hulst1981}. As $p$ increments, $\mathrm{d}
\sigma^{\star} / \mathrm{d} \theta$ exhibits increasingly sharper
features in frequency and angle due to the oscillations of $S(q)$,
giving considerably different weights to the forward and backward
scattering. Clearly, for low $p$ the scattering is primarily
forward, whereas for strongly-correlated disorder the forward
scattering is inhibited in a broad range of frequencies.

This redistribution of the scattered light is at the core of the
modification of the transport properties in correlated
disorder. To illustrate this point, we calculate the transport
mean free path $\ell_t$ in the correlated system~\footnote{Note that
the expression used in Refs.~\cite{Yang2010,Noh2011} applies
exclusively to 3D systems.}
\begin{equation}\label{eq3}
\ell_t=\left( \rho \int_{0}^{\pi} \frac{\mathrm{d}
\sigma^{\star}}{\mathrm{d} \theta} (1-\cos \theta)
\mathrm{d}\theta \right)^{-1},
\end{equation}
where $\rho$ is the number density of scatterers, and the scattering anisotropy factor
$g$
\begin{equation}\label{eq4}
g=\frac{1}{\sigma^{\star}} \int_{0}^{\pi}
\frac{\mathrm{d}\sigma^{\star}}{\mathrm{d} \theta} \cos \theta
\mathrm{d}\theta,
\end{equation}
which indicates the degree of anisotropy of the effective single
scattering event. The results are shown in Fig.~\ref{fig2}(b-c) as
a function of $p$. First, as expected, correlations yield spectral
ranges with either longer or shorter transport mean free paths,
the latter occurring in particular when
$\lambda_e=2nd_{\text{avg}}$, with $n=1,2,...$ (Bragg-like
scattering) and $d_{\text{avg}}$ is the average distance between
nearest-neighbour scatterers. Variations larger than a factor of 2
are observed. Second, the anisotropy factor $g$ for highly
correlated structures becomes negative on broad frequency ranges,
reaching values as low as -0.9, indicating a strong backward
scattering~\cite{RojasOchoa2004}. Interestingly, this leads to a
peculiar light transport in which the scattering mean free path
$\ell_s=\ell_t(1-g)$ is longer than the transport mean free path
$\ell_t$. This scattering property is rare in systems of isolated
particles, and it has been observed only in specific
cases~\cite{Gomez-Medina2012}.

Gaining control over the transport mean free path provides an
unprecedented control on light localization phenomena. In this
respect, two-dimensional structures are very peculiar since the
dependence of the localization length on the transport mean free
path is critical, so that a small change in $\ell_t$ should yield
dramatic changes of $\xi$. The localization length is indeed
predicted to be given by~\cite{Sheng2010}:
\begin{equation}\label{eq5}
\xi \approx \ell_t \exp \left[\pi^2 \frac{\ell_t}{\lambda_e} \right].
\end{equation}
We therefore expect that a modification of the degree of
correlation $p$ could lead to modifications of the localization
length $\xi$ over orders of magnitude, making it possible to go
from a quasi-extended to a localized regime easily in finite-size
systems. In Fig 2(d), the localization length $\xi$ predicted
semi-analytically is shown in semi-log scale. The strong
modulation of $\xi$ as a function of frequencies is striking. In
the 2D correlated system, one goes from a regime in which $\xi$ is
much larger than any realistic system (low frequencies) to a
regime in which the two can be comparable (at $a/\lambda \approx
0.2$). Although we do not expect Eq.~\ref{eq5} to be
quantitatively accurate~\footnote{Eq.~\ref{eq5} is derived from a
renormalization of the diffusion constant in which it is assumed
that the correction term is small~\cite{Sheng2010}. Also, $\ell_t$
is obtained neglecting recurrent scattering and near-field
interactions. Since minute variations of $\ell_t$ leads to large
variations in $\xi$, we do not expect to have a quantitatively
accurate estimation of $\xi$ for our dense systems.}, this huge
photonic dispersion (variation of orders of magnitude within
$\Delta \omega/\omega_0 =0.2$), suggests that we could truly
observe a dramatic variation of $\xi$ in real systems.

\begin{figure}[h!]
   \includegraphics[width=7cm]{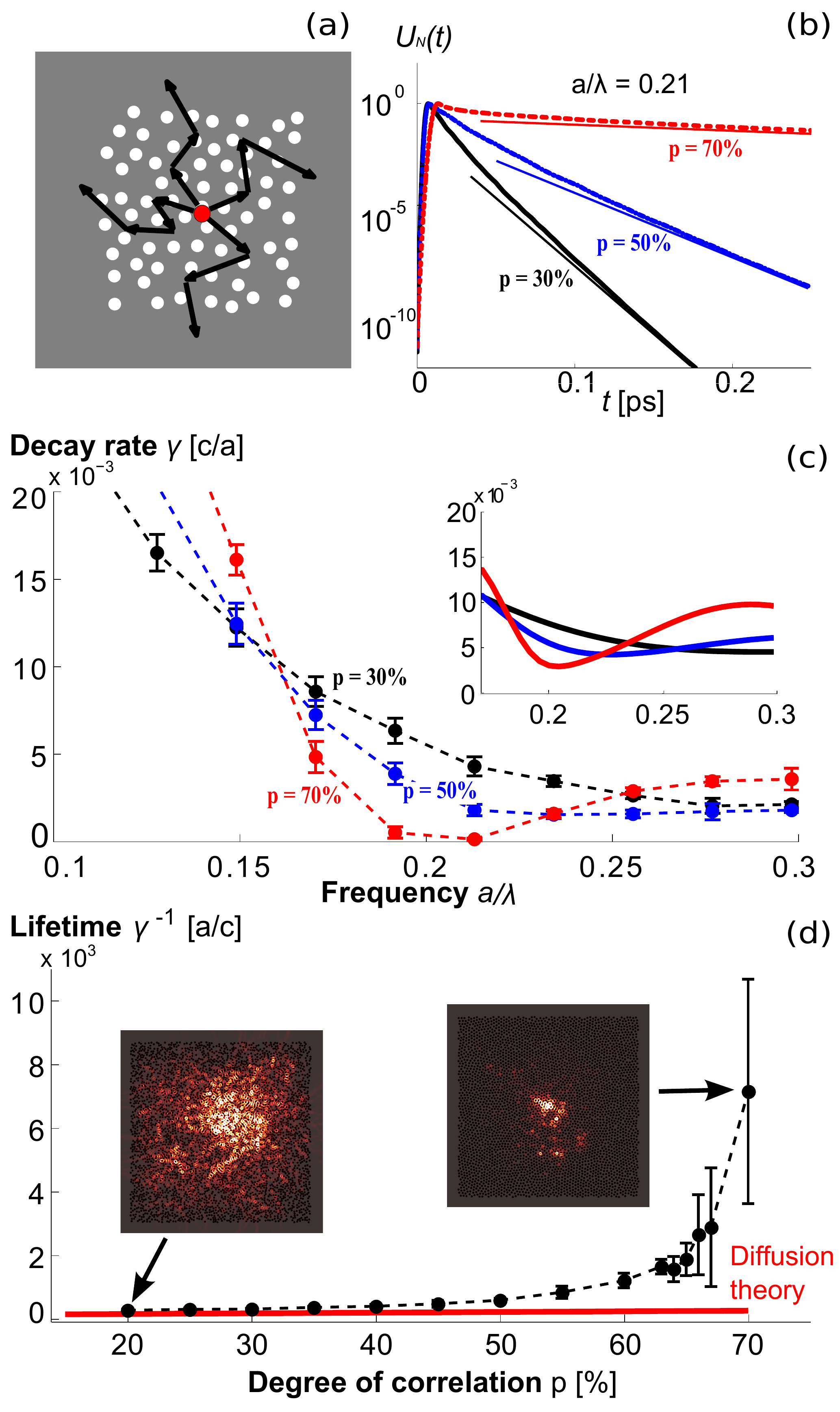}
\caption{\textbf{Numerical FDTD simulations} -- (a) Sketch of the
simulated system for the evaluation of the average decay constant
of the modes. (b) Normalized integrated energy density $U_n(t)$
versus time at frequency $a/\lambda=0.21$, for different $p$. The
straight lines are decaying exponentials, emphasizing the
multi-exponential decay observed at short times. The decay
constant values $\gamma$ were estimated at long times. (c)
Numerically estimated decay constants versus frequency as a
function of $p$. The inset shows the decay constants evaluated
semi-analytically from Eq.~\ref{eq6}, only for $L/\ell_t \geq 6$.
(d) Numerically estimated average lifetime $\gamma^{-1}$ as a
function of $p$ at frequency $a/\lambda=0.21$. The red line
represents the prediction according to diffusion theory and the
intensity maps are calculated for different $p$. \label{fig3}}
\end{figure}

To test these predictions, we investigated the transport
properties of the 2D correlated systems through numerical 2D
finite-difference time-domain (FDTD) simulations, using a freely
available software package~\cite{Oskooi2010}. We considered finite
systems with the same structural parameters and open boundaries
(squares of side $L=36a$ surrounded by perfectly matched layers,
see Fig.~\ref{fig3}(a)). The system was excited from a set of
225 randomly distributed dipole sources having an impulse with
bandwidth of 0.02 $\lambda/a$. Since the system is open, the
energy density is expected to decay exponentially at long times as
$U(t)\propto \exp[-\gamma t]$, where $\gamma$ is the decay
constant. A change in the degree of correlation $p$ should yield a
change in $\ell_t$ and thus in the average time needed for light
to escape from the system. This is illustrated in
Fig.~\ref{fig3}(b), where, at a frequency $a/\lambda=0.21$,
increasing $p$ yields a diminution of $\gamma$. The
multi-exponential decay at shorter times is due to the excitation
of several modes in the structure which couple to the environment
with different efficiency. The decay constants  $\gamma$ were
therefore obtained from exponential fits at sufficiently long
times for various frequencies and degrees of correlation $p$, and
an average was performed over 6 disorder realizations. Note that a
proper statistical analysis of $\gamma$ would require more
disorder realizations, which would be extremely time-consuming. We
have observed, however, that 6 disorder realizations are
sufficient to show the increase of the decay constant as the
degree of correlation increases, as reported below
(Fig.~\ref{fig3}(d)).

In Fig.~\ref{fig3}(c), we show the average decay constants of the
photonic structures with $p=30\%$, $50\%$ and $70\%$ as a function
of the pulse excitation frequency, estimated from the numerical
FDTD simulations. The effect of correlations on light transport is
particularly clear. At frequencies close to $a/\lambda \approx
0.2$, $\gamma$ is strongly diminished due to a reduction of
$\ell_t$ and at lower frequencies ($a/\lambda<0.17$) one observes
an increase of $\gamma$, in accordance with the increase of
$\ell_t$ (see Fig.~\ref{fig2}(b)). Note that $\gamma$ drops over 2
orders of magnitude within a relative bandwidth of
$\Delta\omega/\omega_o=0.2$.

It is also interesting to compare the values obtained numerically
with those expected from the semi-analytical approach within the
diffusion approximation. The decay constant for diffuse light in a
2D system open along two directions is given by
\begin{equation}\label{eq6}
\gamma=\frac{2 \pi^2 D}{(L+2 z_e)^2},
\end{equation}
where $z_e=\frac{\pi}{4} \ell_t$ is the so-called extrapolated
length~\cite{vanRossum1999,payne_anderson_2010} (internal
reflections are neglected), and $D=v_e \ell_t / 2$ is the
diffusion constant with $v_e=c/n_e$ the energy velocity. According
to Eq.~\ref{eq3}, the optical thickness $L/\ell_t$ of our systems
can be extremely small at very low frequencies, so that even less
than a single scattering event can occur. In such a regime,
Eq.~\ref{eq6} is not accurate~\cite{Elaloufi2004}. Hence, we limit
our analysis to frequency domains (inset in Fig.~\ref{fig3}(c))
for which $L/\ell_t\geq 6$, so that the accuracy of diffusion
theory is still reasonably good.  While the trends of the decay
constant as a function of frequency and degree of correlation are
in good agreement, only a fair agreement is found quantitatively.
The deviation for $p=70\%$ in particular is marked (more than one
order of magnitude). This discrepancy can be explained by
considering that (i) the expression used to evaluate the transport
mean free path (Eq.~\ref{eq3}) neglects completely recurrent
scattering and near-field phenomena which are likely to occur in
such dense systems (here, the filling fraction is $f=20\%$), and
(ii), more importantly for the strongly correlated system, the
diffusion approximation disregards completely light localization
phenomena, due to interference between multiply-scattered waves,
which are expected to yield a reduction of the diffusion constant.

To better appreciate the occurrence of light localization in these
systems, we retrieve the average lifetime $\gamma^{-1}$ from the
numerical data at frequency $a/\lambda=0.21$ for different $p$, as
shown in Fig.~\ref{fig3}(d). The clear, weakly fluctuating trend
with varying $p$ indicates that 6 disorder realizations already
provide reasonably converged results. The red line represents the
prediction of diffusion theory according to Eqs.~\ref{eq3} and
\ref{eq6}. A clear deviation between numerical calculations and
diffusion theory occurs as $p$ increases. In particular, a
dramatic increase of $\gamma^{-1}$ over an order of magnitude is
observed for $p=70\%$. Such a pronounced effect cannot be
attributed solely to a reduction of  $\ell_t$, since it would be
captured by Eq.~\ref{eq6}, but rather to Anderson localization of
light. Localized modes are, on average, characterized by an
exponentially decaying intensity distribution~\footnote{The
intensity distribution of a localized mode in a single disorder
realization can often deviate strongly from a profile with a clear
exponential decay.}. This implies, due to weak coupling between
these modes and the environment, lifetimes on average that are
much longer than those of quasi-extended modes. Here, the
reduction of $\ell_t$ with increasing $p$ yields a considerable
variation of the localization length $\xi$, which eventually
becomes shorter than the sample size $L$. This is, indeed,
supported by the intensity maps shown in the insets of
Fig.~\ref{fig3}(d), calculated for a single steady-state source
placed in the center of the system. Our interpretation is also
corroborated by the observation of large variations of lifetimes
for different realization of disorder, as shown by the pronounced
errorbars in Fig.~\ref{fig3}d for high $p$, which is expected in
the localized regime. A transition from a quasi-extended regime
($\xi> L$) to a localized regime ($\xi \leq L$) has therefore been
achieved by merely adding short-range correlations in the
disordered system, keeping $f$ and $\o$ unchanged. From these
considerations, we can estimate that $\xi \leq L$, which is 1
order of magnitude smaller with respect to the prediction of
Eq.~\ref{eq5}.

To conclude, we have investigated how short-range correlations
lead to considerable modifications of light transport and
localization phenomena in 2D disordered photonic structures. Using
a semi-analytical approach for the structure factor of the
correlated systems (due to Baus and Colot~\cite{Baus1987}) and a
modified independent scattering
approximation~\cite{Fraden1990,Saulnier1990}, we have investigated
how key transport quantities are affected by short-range
structural correlations. We have found in particular that
short-range correlations make it possible to increase and/or
decrease the localization length by several orders of magnitude.
This shows that it is possible to design structures that are very
weakly scattering and strongly localizing at nearby frequencies.
Two-dimensional disordered systems in which the light transport
and localization is finely-controlled may find interest in the
fundamental study of localization phenomena~\cite{Riboli2011}, the
conception of planar random lasers~\cite{Yang2011}, thin-film
photovoltaic and lighting technologies~\cite{Vynck2012a}, on-chip
random spectrometers~\cite{Redding2013}, or even help to reach the
strong coupling regime with quantum dots or
molecules~\cite{Thyrrestrup2012, Caze2013}.

This work is supported by the Italian National Research Council
(CNR) through the ``EFOR'' project, and the LABEX WIFI (Laboratory
of Excellence within the French Program "Investments for the
Future") under references ANR-10-LABX-24 and ANR-10-IDEX-0001-02
PSL*. The research leading to these results has received funding
from the European Research Council under the European Union's
Seventh Framework Programme (FP7/2007-2013) / ERC grant agreement
number [291349].


\end{document}